\documentclass[fleqn,preprint,showpacs,preprintnumbers,amsmath,amssymb]{revtex4}
\usepackage{graphicx}
\usepackage{dcolumn}
\usepackage{bm}

\begin{document}

\title{Double-wells and double-layers in dusty Fermi-Dirac plasmas: Comparison with the semiclassical Thomas-Fermi counterpart}
\author{M. Akbari-Moghanjoughi}
\affiliation{ Azarbaijan University of
Tarbiat Moallem, Faculty of Sciences,
Department of Physics, 51745-406, Tabriz, Iran, Tel: +(98)411-553-6531}

\date{\today}
\begin{abstract}
\textbf{Based on the quantum hydrodynamic model, a new relationship between the electrostatic-potential and the electron-density in the ultra-dense plasma is derived. Propagation of arbitrary amplitude nonlinear ion waves is, then, investigated in a completely degenerate dense dusty electron-ion plasma, using this new energy relation, for the relativistic electrons in the ground of quantum hydrodynamics model and the results are compared to the case of semiclassical Thomas-Fermi dusty plasma. Using the standard pseudo-potential approach it is remarked that the Fermi-Dirac plasma, in contrast to the Thomas-Fermi counterpart, accommodates a wide variety of nonlinear excitations such as positive-/negative-potential ion solitary and periodic waves, double-layers and double-wells. It is also remarked that, the relativistic degeneracy parameter which relates to the mass-density of plasma has significant effects on the allowed matching-speed range in dusty Fermi-Dirac plasmas.}
\end{abstract}


\maketitle

\section{Background}

The study of plasma under extreme conditions, such as ultrahigh pressure or temperature is growing rapidly and becoming one of the intense fields of plasma research. There are already numerous laboratory applications for plasmas under high pressure such as that found in semiconductors and nanomaterials \cite{haug}. On the other hand, plasmas under high temperature are the main ingredients of inertial confinement fusion and Tokamaks \cite{son1, son2, greaves}. It is known that many astrophysical places are under such extreme conditions \cite{rees, shapiro, misner, miller, michel, goldr}.
The study of ultradense plasmas, however, requires the relativistic and quantum mechanical considerations. For instance, the quantum effects become more important when the Pauli exclusion principal acts effectively on fermions and the interparticle distances get much lower than the characteristic de Broglie thermal wavelength, $h/(2\pi m k_B T)^{1/2}$ \cite{bonitz} which leads to the degeneracy condition. Under a very high external pressure, for instance, the degenerated electrons in matter become relativistic and obey the more general Fermi-Dirac statistics \cite{landau}.

It has shown by S. Chandrasekhar \cite{chandra1} that in a white dwarfs the relativistic pressure, under the influence of gravity, may eventually lead to the collapse of star, due to softening of degeneracy pressure. Using the Fermi-Dirac statistics for electrons, a rigorous mathematical criteria has been developed \cite{chandra2} under which a white dwarf can be considered as completely degenerate and has been shown that a typical white dwarf can be examined successfully under the ideal zero-temperature Fermi-gas model such as that observed in ordinary metals. New study based on semiclassical Thomas-Fermi model reveals that the nonlinear ion-acoustic wave dynamics such a their propagations and collisions in superdense magnetized plasmas are much different when the electrons are ultra-relativistically degenerate from those with non-relativistic electrons \cite{akbari1, akbari2}. Therefore, it would be interesting to explore the effects of relativistic degeneracy degree (relativistic degeneracy parameter) on nonlinear wave dynamics in a white dwarf under high gravitational pressure.

There has been already many investigations on the properties of nonlinear propagations in a degenerate plasma which examine either non-relativistic \cite{sabry, misra1, misra2, chatterjee, akbari3} or ultra-relativistic \cite{rasheed, akbari4, mamun} degeneracy conditions. However, the later studies in ultra-relativistic degeneracy cases are based on a much simplified model of Thomas-Fermi approximation which has its known limitations. In current work, using the quantum hydrodynamics formulation \cite{gardner} and assuming a more appropriate Fermi-Dirac model of degeneracy pressure which accounts for the relativistic effects of electrons, we examine the general case which covers the whole range of the relativistic degeneracy, special limits of which should produce the previous results. The presentation of the article is as follows. The basic normalized plasma equations are introduced in section \ref{equations}. Nonlinear arbitrary-amplitude solutions are derived in section \ref{Sagdeev}. Numerical analysis and discussions is given in section \ref{discussion} and final remarks are presented in section \ref{conclusion}

\section{Basic Fluid Model}\label{equations}

In order to evaluate the nonlinear dynamics of ion waves we use the complete set of quantum hydrodynamics (QHD) equations. Here, we consider a dense plasma consisting of relativistically degenerate cold electrons, dynamic ultra-cold ions and neutralizing static background dust ingredients. It is assumed that such a plasma is collision-less due to Pauli-blocking. It is further required that the electrons obey the zero-temperature Fermi-Dirac statistics. Therefore, the closed set of QHD equation \cite{haas}, in dimensional form, is written as following
\begin{equation}\label{dim}
\begin{array}{l}
\frac{{\partial {n_i}}}{{\partial t}} + \frac{{\partial {n_i}{v_i}}}{{\partial x}} = 0, \\
\frac{{\partial {n_e}}}{{\partial t}} + \frac{{\partial {n_e}{v_e}}}{{\partial x}} = 0, \\
\frac{{\partial {v_i}}}{{\partial t}} + {v_i}\frac{{\partial {v_i}}}{{\partial x}} =  - \frac{e}{{{m_i}}}\frac{{\partial \phi }}{{\partial x}} - \frac{{\gamma {k_B}{T_i}}}{{{m_i n_{i0}}}}\left(\frac{n_i}{n_{i0}}\right)^{\gamma  - 2}\frac{{\partial {n_i}}}{{\partial x}}, \\
\frac{{{m_e}}}{{{m_i}}}\left( {\frac{{\partial {v_e}}}{{\partial t}} + {v_e}\frac{{\partial {v_e}}}{{\partial x}}} \right) = \frac{e}{{{m_i}}}\frac{{\partial \phi }}{{\partial x}} - \frac{1}{{{m_i}{n_e}}}\frac{{\partial {P_e}}}{{\partial x}} + \frac{{{\hbar ^2}}}{{2{m_e}{m_i}}}\frac{\partial }{{\partial x}}\left( {\frac{1}{{\sqrt {{n_e}} }}\frac{{{\partial ^2}\sqrt {{n_e}} }}{{\partial {x^2}}}} \right), \\ \frac{{{\partial ^2}\phi }}{{\partial {x^2}}} = 4\pi e({n_e} - {n_i} + {z_d}{n_d}), \\
\end{array}
\end{equation}
where, $n_j$, $m_j$, $z_j$ and $v_j$ denote the $j$-th species number-density, mass, atomic-number and velocities with $j=\{e,i,d\}$ for electron, ion, dust, respectively. Note that the quantity $v_e$ stands for the relativistic electron speed and $n_{i0}$ denotes the ion equilibrium density. Also $\gamma$, $T_i$ and $\hbar$, respectively, present the adiabatic constant, ion-temperature and the scaled Plank-constant. Furthermore, the following scalings can be used in order to obtain the normalized set of equations
\begin{equation}\label{T}
x \to \frac{{{c_s}}}{{{\omega _{pi}}}}\bar x,\hspace{3mm}t \to \frac{{\bar t}}{{{\omega _{pi}}}},\hspace{3mm}n_{e,i} \to \bar n_{e,i}{n_0},\hspace{3mm}v_{e,i} \to \bar v_{e,i}{c_s},\hspace{3mm}\phi  \to \bar \phi \frac{{{m_e}{c^2}}}{e},
\end{equation}
where, the bar symbols denote the dimensionless quantities which are omitted in forthcoming algebra for simplicity. The parameters ${\omega _{pi}} = \sqrt {4\pi {e^2}{n_{i0}}/{m_i}}$ and ${c_{s}} = \sqrt {{m_e}{c^2}/{m_i }}$ are characteristic plasma frequency and sound-speed, respectively, and $c$ is the speed of light in vacuum. Also, $n_{0}$ denotes the normalizing number-density, ${n_0} = \frac{{8\pi m_e^3{c^3}}}{{3{h^3}}}\simeq 5.9 \times 10^{29} cm^{-3}$ a value in white dwarf electron number-density range. The electron relativistic degeneracy pressure in zero-temperature configuration, which governs most dense astrophysical objects such as white dwarfs, can be expressed in the following general form \cite{chandra1}
\begin{equation}
{P_e} = \frac{{\pi m_e ^4{c^5}}}{{3{h^3}}}\left\{ {{R}\left( {2{R^2} - 3} \right)\sqrt {1 + {R^2}}  + 3\ln \left[ {R + \sqrt {1 + {R^2}} } \right]} \right\},
\end{equation}
in which $R=(n_e/n_0)^{1/3}$ (${n_0} = \frac{{8\pi m_e^3{c^3}}}{{3{h^3}}}\simeq 5.9 \times 10^{29} cm^{-3}$) is the relativistic degeneracy parameter measuring the relative Fermi-momentum of electrons, i.e. $R=p_{Fe}/m_e c$. In the limits of very small and very large values of the relativity parameter, $R$, we obtain the well known non- and ultra-relativistic degeneracy pressures (e.g. see Ref. \cite{chandra1})
\begin{equation}\label{limits}
P_e = \left\{ {\begin{array}{*{20}{c}}
{\frac{1}{{20}}{{\left( {\frac{3}{\pi }} \right)}^{\frac{2}{3}}}\frac{{{h^2}}}{{{m_e}}}n_e^{\frac{5}{3}}\hspace{10mm}(R \to 0)}  \\
{\frac{1}{{8}}{{\left( {\frac{3}{\pi }} \right)}^{\frac{1}{3}}}hcn_e^{\frac{4}{3}}\hspace{10mm}(R \to \infty )}  \\
\end{array}} \right\}.
\end{equation}
Using the scalings introduced above, the normalized hydrodynamics set of basic equations, including the quantum force effect, is written as
\begin{equation}\label{normal}
\begin{array}{l}
\frac{{\partial {n_i}}}{{\partial t}} + \frac{{\partial {n_i}{v_i}}}{{\partial x}} = 0, \\
\frac{{\partial {n_e}}}{{\partial t}} + \frac{{\partial {n_e}{v_e}}}{{\partial x}} = 0, \\
\frac{{\partial {v_i}}}{{\partial t}} + {v_i}\frac{{\partial {v_i}}}{{\partial x}} =  - \frac{{\partial \phi }}{{\partial x}} - \left( {\frac{{\gamma {k_B}{T_i}}}{{n_{i0}{m_e}{c^2}}}} \right)\left(\frac{n_i}{n_{i0}}\right)^{\gamma  - 2}\frac{{\partial {n_i}}}{{\partial x}}, \\
\frac{{{m_e}}}{{{m_i}}}\left( {\frac{{\partial {v_e}}}{{\partial t}} + {v_e}\frac{{\partial {v_e}}}{{\partial x}}} \right) = \frac{{\partial \phi }}{{\partial x}} - \frac{{\partial  }}{{\partial x}} \sqrt {1 + {R_0^{2} {\alpha^{2/3}}{n_e}^{2/3}}} + \frac{{{H_r^{2}}}}{2}\left( {\frac{{{m_e}}}{{{m_i}}}} \right)\frac{\partial }{{\partial x}}\left( {\frac{1}{{\sqrt {{n_e}} }}\frac{{{\partial ^2}\sqrt {{n_e}} }}{{\partial {x^2}}}} \right), \\
\frac{{{\partial ^2}\phi }}{{\partial {x^2}}} = {n_e} - {n_i} + {z_d}{n_d}, \\
\end{array}
\end{equation}
where, $R_0=(n_{i0}/n_0)^{1/3}$ is a measure of the relativistic degeneracy effects (normalized relativity parameter), ${\alpha}=n_{e0}/n_{i0}$ and $(m_e/m_i)H_r^{2}/2=\hbar^2 \omega_{pi}^2/(m_e c^2)^2$, is the quantum diffraction coefficient which is related to the relativity parameter via $(m_e/m_i)H_r^{2}/2\simeq 4.6\times 10^{-10} R_0^{3}$. It is clearly confirmed that the quantum correction in this model is negligible even for higher values of relativity parameter. For instance, for the value of $R_0=1000$ we obtain a quantum correction of order $\simeq 0.46$. It should also be noted that the relativity parameter, $R_0$, (using $\rho\simeq 2m_p(n_{i0}+{z_d}{n_d})=2m_p n_{i0}(2-\alpha)$ with $m_p$ the proton mass) is related to the mass density through $\rho \simeq 3.94\times 10^6 (2-\alpha)R_0^{3}(gr/cm^{3})$. Taking, for instance, the density of typical white dwarfs to be in the range $10^5<\rho(gr/cm^{3})<10^{8}$, results in typical values of $0.37<R_0<8$ for the relativity parameter. However, the current study does not limit the applicability of results to specific density range. Now, neglecting the small contributing terms containing the ratio $m_e/m_i$ and assuming that $k_B T_i \ll m_e c^2$, we are led to the following simplified set of normalized equations
\begin{equation}\label{comp}
\begin{array}{l}
\frac{{\partial {n_i}}}{{\partial t}} + \frac{{\partial {n_i}{v_i}}}{{\partial x}} = 0, \\
\frac{{\partial {n_e}}}{{\partial t}} + \frac{{\partial {n_e}{v_e}}}{{\partial x}} = 0, \\
\frac{{\partial {v_i}}}{{\partial t}} + {v_i}\frac{{\partial {v_i}}}{{\partial x}} =  - \frac{{\partial \phi }}{{\partial x}}, \\
\frac{{\partial \phi }}{{\partial x}} - \frac{{\partial}}{{\partial x}} \sqrt {1 + {R_0^{2}  {\alpha^{2/3}}{n_e}^{2/3}}} =0, \\
\frac{{{\partial ^2}\phi }}{{\partial {x^2}}} = {n_e} - {n_i} + (1-\alpha). \\
\end{array}
\end{equation}
Note, however, that the overall neutrality condition, $(z_d n_{d0})/n_{i0}=(1-\alpha)$, is met at $\phi=0$. Solving the basic equations (Eqs. (\ref{comp})) for the electron and ion number-densities in terms of electrostatic-potential, using the quasi-neutrality condition, and employing the appropriate boundary requirement ($\mathop {\lim }\limits_{{\phi} \to 0} {n_{e,i}} = \{\alpha,1\}$), results in the following energy relations
\begin{equation}\label{phis}
\begin{array}{l}
{n_i} = \frac{1}{{\sqrt {1 - \frac{{2\phi }}{{{M^2}}}} }}, \\
{n_e} = {\alpha ^{ - 1}}R_0^{ - 3}{\left[ {R_0^{2}{\alpha ^{4/3}} + \phi \left( {2\sqrt {1 + R_0^{2}{\alpha ^{4/3}}}  + \phi } \right)} \right]^{3/2}}.\\
\end{array}
\end{equation}
\textbf{The energy relation for electrons resembles to that of known Thomas-Fermi semiclassical model ($n_e=\alpha(1+\phi)^{3/2}$) used in some literature \cite{abdelsalam, akbari1, akbari2, akbari4}. However, this new expression contains more information about the electron dynamics as it contains an extra parameter which describes the relativistic dynamical properties of electrons (electron relativistic Fermi-momentum , i.e. $R=p_{Fe}/m_e c$) in an ultra-dense plasma. It should be noted that the Thomas-Fermi approximation is valid only for slow variation of electrostatic potential and its application is limited for the ultradense plasmas where electron exchange potential can be dominant and one should appeal for the Thomas-Fermi-Dirac approximation \cite{levine}. An advantage of this model with respect to Thomas-Fermi is that the current model accounts for the whole range of relativity parameter the two limits of which leads to the well known non- and ultra relativistic degeneracy pressure limits given in Eq. (\ref{limits}).}

\textbf{In the proceeding section we aim at finding the appropriate pseudo-potential which describes the propagation of nonlinear ion-acoustic waves in the plasma described by Eq. (\ref{comp}) and compare the results with that obtained in the Thomas-Fermi model.}

\section{Arbitrary-amplitude Analysis}\label{Sagdeev}

In this section we aim at finding the appropriate Sagdeev potential in plasma defined by Eq. (\ref{normal}) which satisfies the nonlinear ion-solitary (ISWs) as well as periodic (IPWs) solutions. To do this using the standard procedure, we transform into a moving coordinate relative to the wave-train. \textbf{Thus, using the new coordinate $\xi=x-Mt$ in order to find the stationary solution with $M$ being the matching-speed of stationary frame not to be confused by the Mach number (e.g. see Ref. \cite{dubinov}), a measure of the wave-speed relative to the normalizing parameter, $c_s$,} and substituting Eqs. (\ref{phis}) into Poisson's equation, multiplying by $\frac{{d\phi }}{{d\xi }}$ and integrating once with suitable boundary conditions mentioned in the previous section, we derive the well-known energy integral
\begin{equation}\label{energy}
\frac{1}{2}{\left( {\frac{{d\phi }}{{d\xi }}} \right)^2} + U(\phi ) = 0,
\end{equation}
where, the desired Sagdeev potentials $U(\phi)$ for the Fermi-Dirac model, read as
\begin{equation}\label{S1}
\begin{array}{l}
{U_{FD}}(\phi ) = - 8\alpha {R_0^{3}}\left[ {(1 - \alpha )\phi  + M\sqrt {{M^2} - 2\phi } } \right] + R_0{\alpha ^{2/3}}\left( {2{R_0^{2}}{\alpha ^{4/3}} - 3} \right)\sqrt {1 + {R_0^{2}}{\alpha ^{4/3}}}  \\
- \left[ {3\left( {2{\phi ^2} - 1} \right)\sqrt {1 + {R_0^{2}}{\alpha ^{4/3}}}  + \phi \left( {2{\phi ^2} + 1} \right) + 2{R_0^{2}}{\alpha ^{4/3}}\left( {3\phi  + \sqrt {1 + {R_0^{2}}{\alpha ^{4/3}}} } \right)} \right] \times  \\
\times \sqrt {{R_0^{2}}{\alpha ^{4/3}} + \phi \left( {\phi  + 2\sqrt {1 + {R_0^{2}}{\alpha ^{4/3}}} } \right)}  + 3{\sinh ^{ - 1}}\left( {R_0{\alpha ^{2/3}}} \right) + 3\ln 2 + 8\alpha {M^2}R_0^{3} \\
- 3\ln \left[ {2\left( {\phi  + \sqrt {1 + {R_0^{2}}{\alpha ^{4/3}}} } \right) + 2\sqrt {{R_0^{2}}{\alpha ^{4/3}} + \phi \left( {\phi  + 2\sqrt {1 + {R_0^{2}}{\alpha ^{4/3}}} } \right)} } \right]. \\
\end{array}
\end{equation}
\textbf{Similarly, for the Thomas-Fermi plasma of this kind, we obtain}
\begin{equation}\label{S2}
{U_{TF}}(\phi ) = {M^2} + \frac{{2\alpha }}{5} - \phi  + \alpha \phi  - \frac{2}{5}\alpha {(1 + \phi )^{5/2}} - {M^2}\sqrt {1 - \frac{{2\phi }}{{{M^2}}}}.
\end{equation}
\textbf{It is remarked that the values of $U(\phi)$ and ${\partial _\phi }U(\phi )$ both vanish at $\phi=0$ for both pseudo-potentials, as expected. From Eqs. (\ref{phis}) we may also derive the maximum values of $\phi$ in terms of other plasma parameters for Fermi-Dirac case, as
\begin{equation}\label{phim1}
{\phi _{m1}} = \frac{{{M^2}}}{2},
\end{equation}
\begin{equation}\label{phim3}
{\phi _{m2}} = \pm 1 - \sqrt {1 + {R_0^{2}}{\alpha ^{4/3}}},
\end{equation}
and one maximum and minimum values of $\phi$ for the Thomas-Fermi case, as
\begin{equation}\label{phim1}
{\phi _{max}} = \frac{{{M^2}}}{2},\hspace{3mm}{\phi _{min}} = -1,
\end{equation}}
The possibility of solitary waves then requires the following conditions to met, simultaneously
\begin{equation}\label{conditions}
{\left. {U(\phi)} \right|_{\phi = 0}} = {\left. {\frac{{dU(\phi)}}{{d\phi}}} \right|_{\phi = 0}} = 0,{\left. {\frac{{{d^2}U(\phi)}}{{d{\phi^2}}}} \right|_{\phi = 0}} < 0.
\end{equation}
It is further required that for at least one either maximum or minimum nonzero $\phi$-value, we have $U(\phi_{m})=0$, so that for every value of $\phi$ in the range ${\phi _m} > \phi  > 0$ or ${\phi _m} < \phi  < 0$, $U(\phi)$ is negative. In such a condition we can obtain a potential minimum which describes the possibility of ISWs propagation.

On the other hand, the existence of a nonlinear periodic wave requires that
\begin{equation}\label{conditions2}
{\left. {U(\phi)} \right|_{\phi = 0}} = {\left. {\frac{{dU(\phi)}}{{d\phi}}} \right|_{\phi = 0}} = 0,{\left. {\frac{{{d^2}U(\phi)}}{{d{\phi^2}}}} \right|_{\phi = 0}} > 0,
\end{equation}
It is also required that for at least one either maximum or minimum nonzero $\phi$-value, we have $U(\phi_{m})=0$, so that for every value of $\phi$ in the range ${\phi _m} > \phi  > 0$ or ${\phi _m} < \phi  < 0$, $U(\phi)$ is positive. In such a condition we can obtain a potential maximum which describes the possibility of IPWs propagation.

Since, the Sagdeev potential has triple roots, there is another possibility of obtaining double-layer (DLs) or double-wells (DWs). The condition for a solitary (periodic) DLs is ${\left. {U(\phi )} \right|_{\phi  = {\phi _m}}} = {\left. {dU(\phi )/d\phi } \right|_{\phi  = {\phi _m}}} = 0$ and ${\left. {{d^2}U(\phi )/d{\phi ^2}} \right|_{\phi  = {\phi _m}}} < 0$ (${\left. {{d^2}U(\phi )/d{\phi ^2}} \right|_{\phi  = {\phi _m}}} > 0$). Existence of a DW, on the other hand, requires that $U({\phi _{m1}}) = U({\phi _{m2}}) = 0$, where, $\phi _{m1}$ and $\phi _{m2}$ are oppositely signed. A DW is condition in which rarefactive ($\phi_m<0$) and compressive ($\phi_m>0$) nonlinear excitations coexist. Since, evaluation of the complete set of conditions on the pseudo-potential analytically is not possible in current case, in the proceeding section a numerical scheme will be employed.

\section{Discussion}\label{discussion}

Detailed numerical investigation of the Sagdeev potential Eq. (\ref{S1}) indicates that a wide variety of nonlinear excitations can be present in a dusty Fermi-Dirac plasma define in Eq. (\ref{normal}). Figure 1 shows positive and negative potential arbitrary amplitude ISWs and IPWs and the variation of the potential profiles with respect to relative electron concentration, $\alpha$, for fixed values of matching-speed, $M$, and relativity parameter, $R_0$ (or plasma mass-density). It is remarked that the increase/decrease in fractional electron concentration (decrease/increase in relative dust concentration) for rarefactive ISWs leads to increase/decrease in soliton amplitude/width, while for IPWs this increase has only minor effects (increase/decreases) on the width/amplitude of the soliton. For IPWs, on the other hand, the increase of relative electron number-density leads to the decrease/increase of wave amplitude/width for both positive and negative potential ($\phi_m<0$) waves.

Figure 2 shows the formation of W-shaped positive and negative potential periodic-DLs (Figs. 2(a) and 2(b)) and solitary-DWs (Figs. 2(c) and 2(d)) in a Fermi-Dirac dusty plasma. From Fig. 2(c) it is clearly observed that the compressive soliton shapes are affected via the variation of the matching-speed more than that of rarefactive ones. On the other hand, Fig. 2(d) reveals that the variation of relative electron concentration effects the rarefactive solitary excitation profiles more than that for compressive ones. It is also noted that the increase of $\alpha$ on rarefactive and compressive solitary profiles, discussed in previous paragraph are quite similar to those of solitary-DWs.

\textbf{In Fig. 3 we showed the variation allowed matching-speed range for ISWs with respect to the normalized plasma mass-density, $\bar\rho$, for the case of Fermi-Dirac plasma (using the relation $\bar\rho = \rho/\rho_0 \simeq (2-\alpha)R_0^{3}$ with $\rho_0 =3.94\times 10^6(gr/cm^{3})$) which is obviously not possible for the case of the Thomas-Fermi plasma. It is remarked that, the speed range of possible propagation of ISWs increases as the normalized plasma mass-density increases for all values of the relative electron concentration. Note that the numerical values of $c_s$ is much higher (compared to the electron Fermi-Speed) in here compared to that in ordinary non-degenerate plasmas. It is also revealed from Figs. 3(a)-3(d) that the increases of the relative electron concentration in the Fermi-Dirac plasma leads to the increases in the allowed range of the matching speed.}

The regions of possibility of these nonlinear excitation are shown in colors in Fig. 4 with each color representing a specific type of nonlinear excitation in the Fermi-Dirac plasma. There is also a plot (Fig 4(d)) with numerated regions for non-color option. The description of different region for nonlinear excitations is as follows
\\
Region-1 (in blue color) indicates the possibility of positive potential IPWs an example of which is shown in Fig. 1(c).\\
Region-2 (in black color) indicates the possibility of solitary-DWs an example of which is shown in Fig. 2(c) and 2(d).\\
Region-3 (in red color) indicates the possibility of negative potential IPWs an example of which is shown in Fig. 1(d).\\
Region-4 (in yellow color) indicates the possibility of compressive ISWs an example of which is shown in Fig. 1(b).\\
Region-3 (in green color) indicates the possibility of rarefactive ISWs an example of which is shown in Fig. 1(a).\\

The periodic-Dls, examples of which is shown in Figs. 2(a) and 2(b), exist at the left borders of regions 1 (yellow), 2 (black) and 3 (red), where, at the left borders of regions 1 (yellow) and 2 (black) there exist positive potential-DLs and the negative potential-DLs occur at the left border of region 3 (red). It is further remarked from Fig. 3 that, there is a region for approximately $3.75>\alpha>0.05$ for which no IPWs are present and that this range is affected only slightly with the variation of the relativistic degeneracy parameter. However, as the parameter $R_0$ which represents the mass-density of the plasma increases the regions expand nearly in uniform manner and move to higher Mach-number values.

\textbf{Finally, in Fig. 5, we show the regions of existence for ISWs (region in yellow labeled 4) and IPWs (region in red labeled 3) for the corresponding dusty plasma in Thomas-Fermi model. It is clearly remarked that, only two regions, namely compressive ISWs and negative potential IPWs regions, are present in the case Thomas-Fermi dusty plasma.}

\section{Concluding Remarks}\label{conclusion}

The Sagdeev pseudo-potential approach was used to investigate the propagation of nonlinear ion excitations in a dusty Fermi-Dirac electron-ion plasma with considering the effect of electron relativistic degeneracy. The matching criteria of existence of such solitary excitations were numerically investigated in terms of the relativity parameter, $R_0$, the relative dust concentration, $(1-\alpha)$, and the allowed matching-speed, $M$ and there was found five distinct regions which accommodate a variety of nonlinear excitations. Nonlinear double-wells and double-layers are also found to exist in such plasma. Furthermore, it was confirmed that the relativistic degeneracy parameter has significant effect on the range of allowed matching-speed in dusty Fermi-Dirac plasmas. The current findings can be applicable to highly degenerate dense astrophysical compact objects such as white dwarfs.

\newpage

\textbf{FIGURE CAPTIONS}

\bigskip

Figure-1

\bigskip

(Color online) Seudo-potential profiles in Fermi-Dirac dusty plasma and their variations with respect to the fractional electron concentration, $\alpha$, for fixed other plasma parameters. The dash size is correspondingly varied with the change in the value of $\alpha$ parameter.

\bigskip

Figure-2

\bigskip

(Color online) The formation of positive-/negative-potential double-layers (DLs) and double-wells (DWs) in Fermi-Dirac dusty plasma. The dash size is correspondingly varied with the change in the value of $\alpha$ or $M$ parameters.

\bigskip

Figure-3

\bigskip

(Color online) The variations of allowed ISW matching-speed with respect to normalized plasma mass-density, $\bar\rho$, and the relative electron concentration, $\alpha$ in the Fermi-Dirac dusty plasma.

\bigskip

Figure-4

\bigskip

(Color online) Regions of existence of different nonlinear excitations shown in colors (blue-black-red-yellow-green) enumerated, respectively, (1-2-3-4) in Fig. 3(d), in Fermi-Dirac dusty plasma. Each region belong to a distinct kind of nonlinear wave (refer to text). Variations in the regions with respect to the relativistic degeneracy parameter, $R_0$, which also reflects the plasma mass-density through the relation, $\rho \simeq 3.97 \times 10^{6}(2-\alpha)R_0^{3} (gr/cc)$, is shown through plots (a)-(d).

\bigskip

Figure-5

\bigskip

(Color online) (a) Regions of existence for ISWs (region in yellow labeled 4) and IPWs (region in red labeled 3) and (b) the variation of pseudo-potential for ISWs with respect to the change in relative electron concentration for the fixed matching-speed, in a Thomas-Fermi dusty plasma.

\newpage

\begin{figure}
\resizebox{1\columnwidth}{!}{\includegraphics{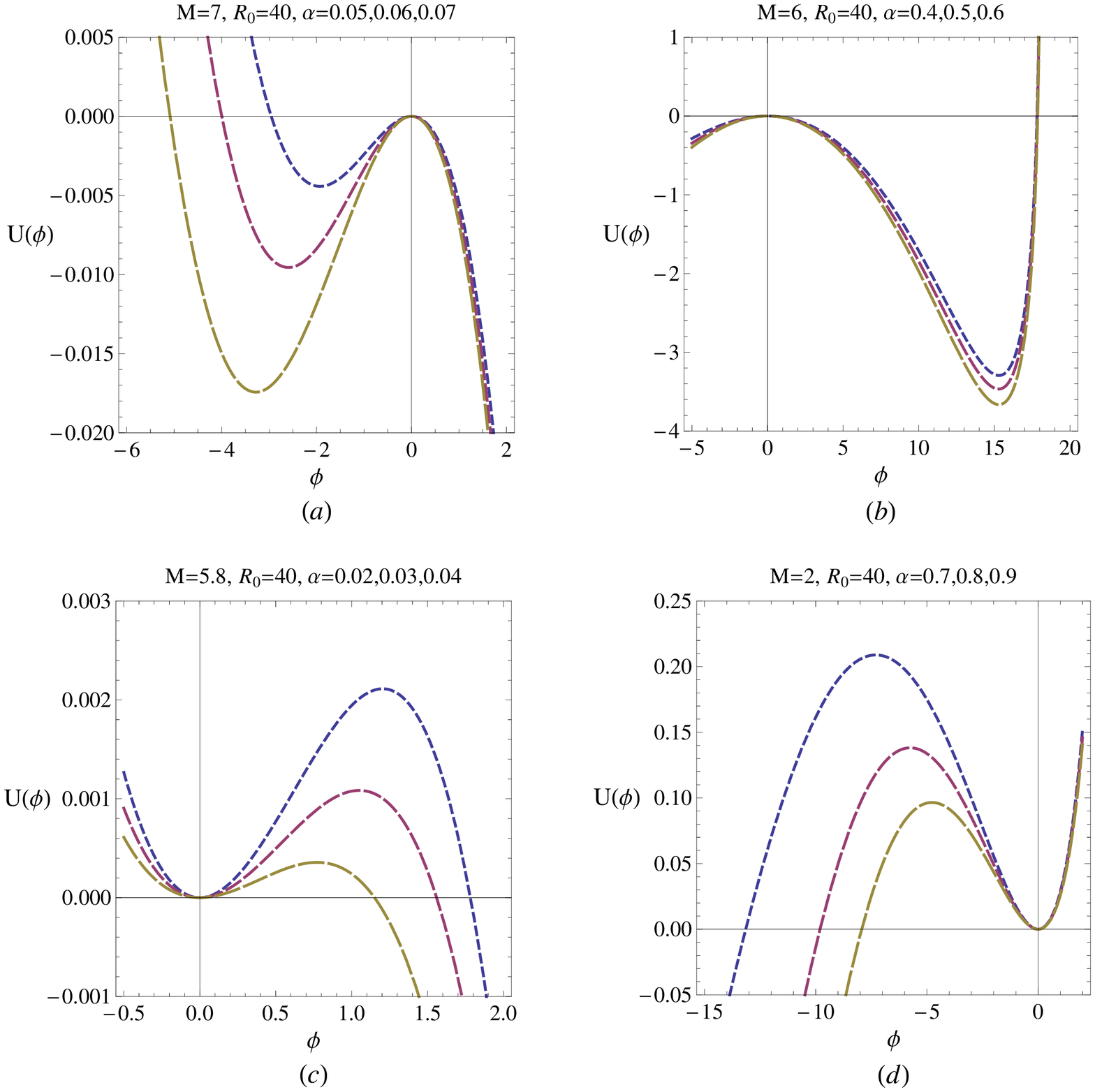}}
\caption{}
\label{fig:1}
\end{figure}

\newpage

\begin{figure}
\resizebox{1\columnwidth}{!}{\includegraphics{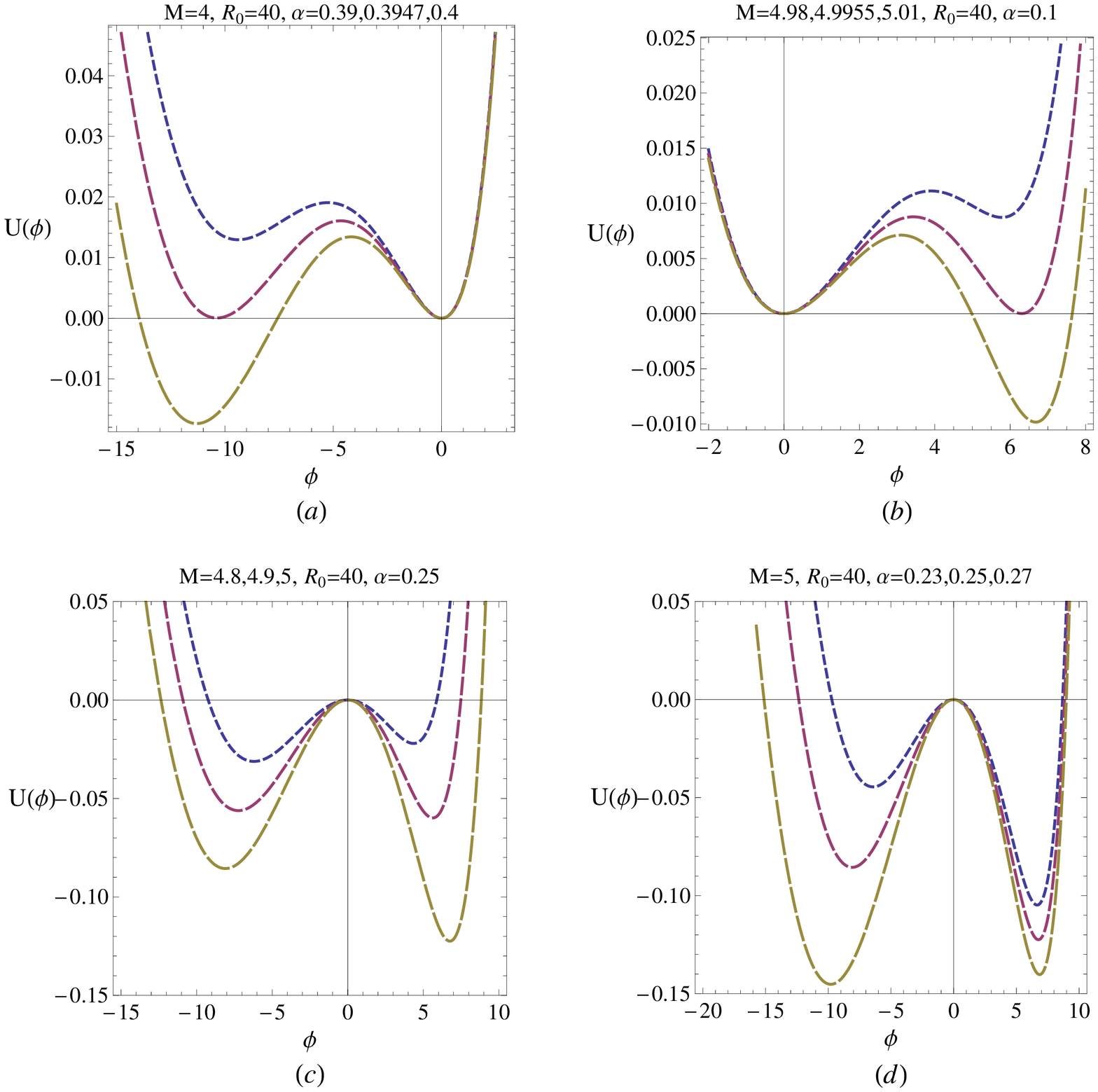}}
\caption{}
\label{fig:2}
\end{figure}

\newpage

\begin{figure}
\resizebox{1\columnwidth}{!}{\includegraphics{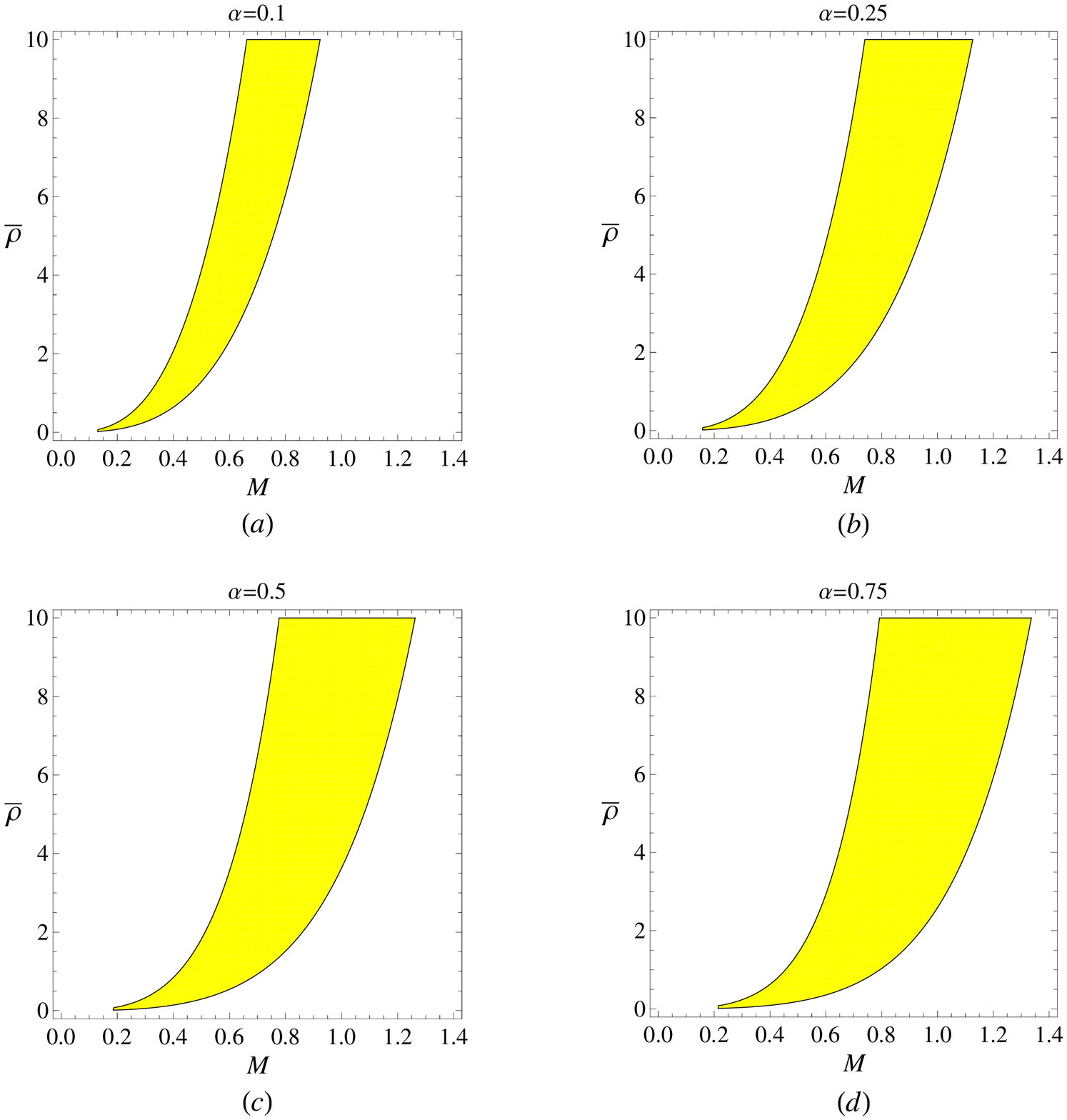}}
\caption{}
\label{fig:3}
\end{figure}

\newpage

\begin{figure}
\resizebox{1\columnwidth}{!}{\includegraphics{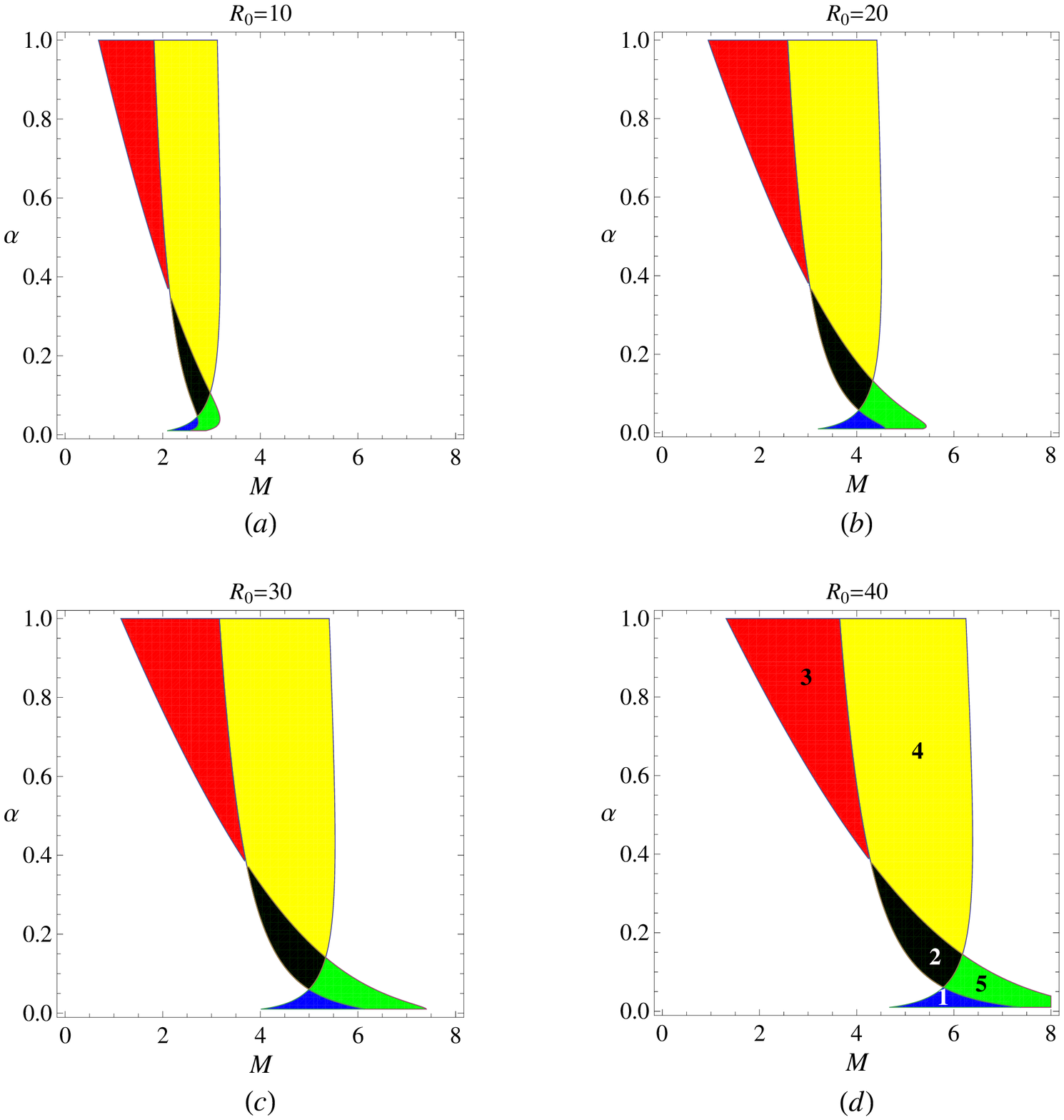}}
\caption{}
\label{fig:4}
\end{figure}

\newpage

\begin{figure}
\resizebox{1\columnwidth}{!}{\includegraphics{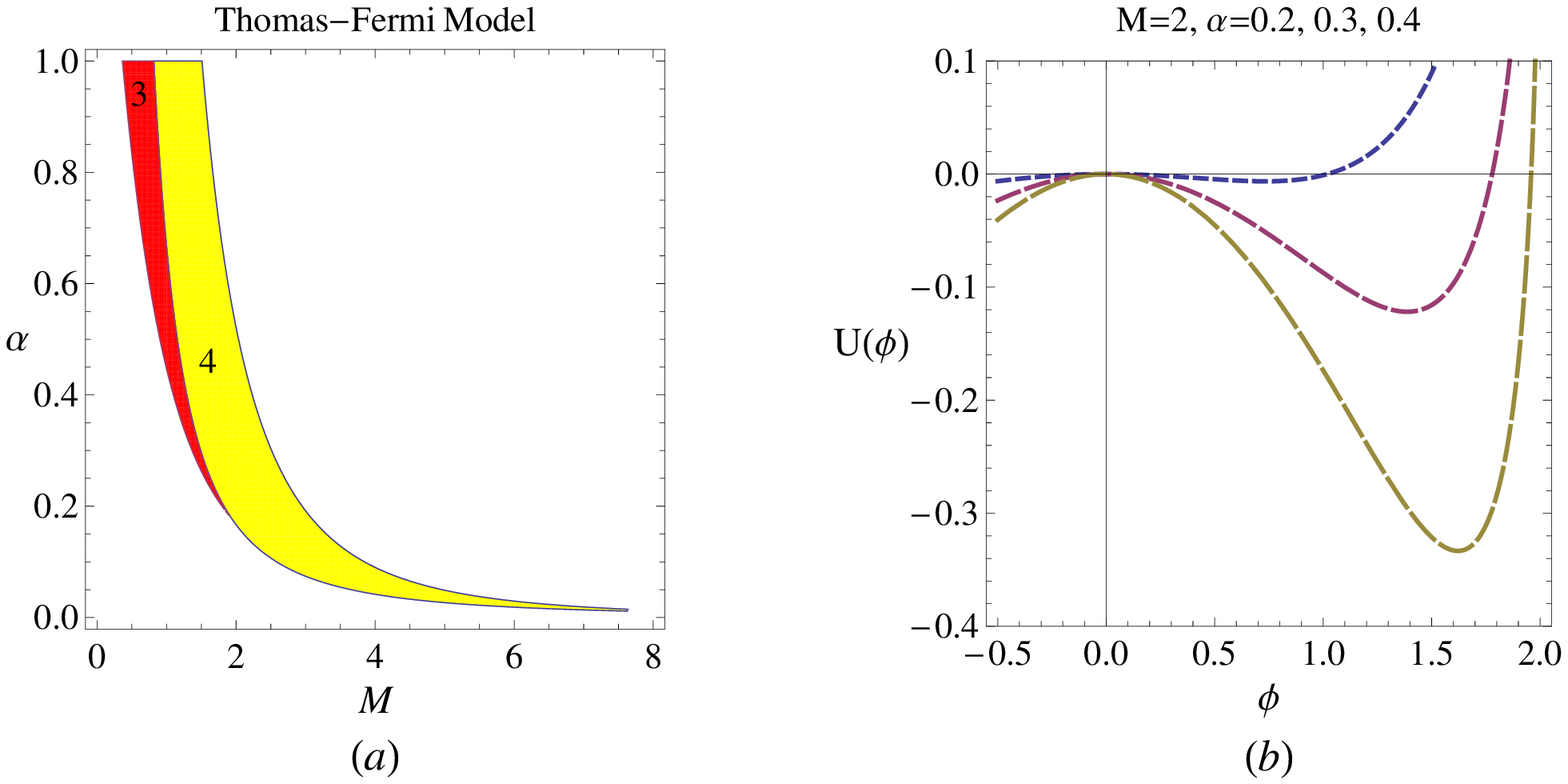}}
\caption{}
\label{fig:5}
\end{figure}

\end{document}